\documentclass[journal]{IEEEtran}
\usepackage{url}
\usepackage[utf8]{inputenc}
\usepackage{xcolor}
\usepackage{amsmath}
\usepackage{amssymb}

\usepackage[acronyms,nonumberlist,nopostdot,nomain,nogroupskip]{glossaries}
\usepackage{tablefootnote}
\usepackage{booktabs}
\usepackage{tabularx}
\usepackage{tikz}
\usepackage{pgfplots}
\pgfplotsset{compat=newest} 
\pgfplotsset{plot coordinates/math parser=false} 
\newlength\fheight
\newlength\fwidth
\usetikzlibrary{plotmarks,patterns,decorations.pathreplacing,backgrounds,calc,arrows,arrows.meta,spy,matrix}
\usepgfplotslibrary{patchplots,groupplots}
\usepackage{tikzscale}
\usepackage{hyperref}

\usepackage{multirow}
\usepackage{tkz-kiviat}

\usepackage[font=scriptsize]{subcaption}
\usepackage[font=footnotesize]{caption}

\usepackage{mathtools}

\usepackage{dblfloatfix}    
\usepackage{colortbl}

\usepackage{makecell}
\usepackage{diagbox}
\usepackage{tikz-qtree}
\usetikzlibrary{trees} 

\newacronym{3gpp}{3GPP}{3rd Generation Partnership Project}
\newacronym{adc}{ADC}{Analog to Digital Converter}
\newacronym{5g}{5G}{5th generation}
\newacronym{aimd}{AIMD}{Additive Increase Multiplicative Decrease}
\newacronym{am}{AM}{Acknowledged Mode}
\newacronym{amc}{AMC}{Adaptive Modulation and Coding}
\newacronym{aqm}{AQM}{Active Queue Management}
\newacronym{awgn}{AGWN}{Additive White Gaussian Noise}
\newacronym{balia}{BALIA}{Balanced Link Adaptation}
\newacronym{bdp}{BDP}{Bandwidth-Delay Product}
\newacronym{bf}{BF}{beamforming}
\newacronym{cc}{CC}{Congestion Control}
\newacronym{cdf}{CDF}{Cumulative Distribution Function}
\newacronym{cn}{CN}{Core Network}
\newacronym{cqi}{CQI}{Channel Quality Information}
\newacronym{cp}{CP}{Control Plane}
\newacronym{csirs}{CSI-RS}{Channel State Information - Reference Signal}
\newacronym{dc}{DC}{Dual Connectivity}
\newacronym{dce}{DCE}{Direct Code Execution}
\newacronym{dci}{DCI}{Downlink Control Information}
\newacronym{dl}{DL}{Downlink}
\newacronym{dmr}{DMR}{Deadline Miss Ratio}
\newacronym{dmrs}{DMRS}{DeModulation Reference Signal}
\newacronym{e2e}{E2E}{End-to-End}
\newacronym{ecn}{ECN}{Explicit Congestion Notification}
\newacronym{edf}{EDF}{Earliest Deadline First}
\newacronym{enb}{eNB}{evolved Node Base}
\newacronym{epc}{EPC}{Evolved Packet Core}
\newacronym{es}{ES}{Edge Server}
\newacronym{fdma}{FDMA}{Frequency Division Multiple Access}
\newacronym{fdd}{FDD}{Frequency Division Duplexing}
\newacronym[firstplural=Radio Access Technologies (RATs)]{rat}{RAT}{Radio Access Technology}
\newacronym[firstplural=Radio Access Technology (RTs)]{rt}{RT}{Radio Technology}
\newacronym{fs}{FS}{Fast Switching}
\newacronym{ftp}{FTP}{File Transfer Protocol}
\newacronym{gnb}{gNB}{Next Generation Node Base}
\newacronym{harq}{HARQ}{Hybrid Automatic Repeat reQuest}
\newacronym{hetnet}{HetNet}{Heterogeneous Network}
\newacronym{hh}{HH}{Hard Handover}
\newacronym{hol}{HOL}{Head-of-Line}
\newacronym{ia}{IA}{Initial Access}
\newacronym{imt}{IMT}{International Mobile Telecommunication}
\newacronym{iot}{IoT}{Internet of Things}
\newacronym{los}{LOS}{Line of Sight}
\newacronym{lte}{LTE}{Long Term Evolution}
\newacronym{m2m}{M2M}{Machine to Machine}
\newacronym{mac}{MAC}{Medium Access Control}
\newacronym{mc}{MC}{Multi-Connectivity}
\newacronym{mcs}{MCS}{Modulation and Coding Scheme}
\newacronym{mec}{MEC}{Mobile Edge Cloud}
\newacronym{mi}{MI}{Mutual Information}
\newacronym{mimo}{MIMO}{Multiple Input, Multiple Output}
\newacronym{mmwave}{mmWave}{millimeter wave}
\newacronym{mptcp}{MPTCP}{Multipath TCP}
\newacronym{mr}{MR}{Maximum Rate}
\newacronym{mss}{MSS}{Maximum Segment Size}
\newacronym{mtd}{MTD}{Machine-Type Device}
\newacronym{mtu}{MTU}{Maximum Transmission Unit}
\newacronym{nfv}{NFV}{Network Function Virtualization}
\newacronym{nlos}{NLOS}{Non Line of Sight}
\newacronym{nr}{NR}{New Radio}
\newacronym{ofdm}{OFDM}{Orthogonal Frequency Division Multiplexing}
\newacronym{pdcch}{PDCCH}{Physical Downlonk Control Channel}
\newacronym{pdcp}{PDCP}{Packet Data Convergence Protocol}
\newacronym{pdsch}{PDSCH}{Physical Downlink Shared Channel}
\newacronym{pdu}{PDU}{Packet Data Unit}
\newacronym{pf}{PF}{Proportional Fair}
\newacronym{pgw}{PGW}{Packet Gateway}
\newacronym{phy}{PHY}{Physical}
\newacronym{pbch}{PBCH}{Physical Broadcast Channel}
\newacronym[plural=\gls{mme}s,firstplural=Mobility Management Entities (MMEs)]{mme}{MME}{Mobility Management Entity}
\newacronym{prb}{PRB}{Physical Resource Block}
\newacronym{pss}{PSS}{Primary Synchronization Signal}
\newacronym{pucch}{PUCCH}{Physical Uplink Control Channel}
\newacronym{pusch}{PUSCH}{Physical Uplink Shared Channel}
\newacronym{rach}{RACH}{Random Access Channel}
\newacronym{ran}{RAN}{Radio Access Network}
\newacronym{red}{RED}{Random Early Detection}
\newacronym{rf}{RF}{Radio Frequency}
\newacronym{rlc}{RLC}{Radio Link Control}
\newacronym{rlf}{RLF}{Radio Link Failure}
\newacronym{rrc}{RRC}{Radio Resource Control}
\newacronym{rrm}{RRM}{Radio Resource Management}
\newacronym{rr}{RR}{Round Robin}
\newacronym{rs}{RS}{Remote Server}
\newacronym{rsrp}{RSRP}{Reference Signal Received Power}
\newacronym{rss}{RSS}{Received Signal Strength}
\newacronym{rtt}{RTT}{Round Trip Time}
\newacronym{rw}{RW}{Receive Window}
\newacronym{rx}{RX}{Receiver}
\newacronym{sa}{SA}{standalone}
\newacronym{sack}{SACK}{Selective Acknowledgment}
\newacronym{sap}{SAP}{Service Access Point}
\newacronym{sch}{SCH}{Secondary Cell Handover}
\newacronym{scoot}{SCOOT}{Split Cycle Offset Optimization Technique}
\newacronym{sdma}{SDMA}{Spatial Division Multiple Access}
\newacronym{sinr}{SINR}{Signal to Interference plus Noise Ratio}
\newacronym{sm}{SM}{Saturation Mode}
\newacronym{snr}{SNR}{Signal to Noise Ratio}
\newacronym{son}{SON}{Self-Organizing Network}
\newacronym{ss}{SS}{Synchronization Signal}
\newacronym{srs}{SRS}{Sounding Reference Signal}
\newacronym{sss}{SSS}{Secondary Synchronization Signal}
\newacronym{tb}{TB}{Transport Block}
\newacronym{tcp}{TCP}{Transmission Control Protocol}
\newacronym{tdd}{TDD}{Time Division Duplexing}
\newacronym{tdma}{TDMA}{Time Division Multiple Access}
\newacronym{tfl}{TfL}{Transport for London}
\newacronym{tm}{TM}{Transparent Mode}
\newacronym{trp}{TRP}{Transmitter Receiver Pair}
\newacronym{tti}{TTI}{Transmission Time Interval}
\newacronym{ttt}{TTT}{Time-to-Trigger}
\newacronym{tx}{TX}{Transmitter}
\newacronym{ue}{UE}{User Equipment}
\newacronym{ul}{UL}{Uplink}
\newacronym{uml}{UML}{Unified Modeling Language}
\newacronym{um}{UM}{Unacknowledged Mode}
\newacronym{utc}{UTC}{Urban Traffic Control}
\newacronym{vm}{VM}{Virtual Machine}
\newacronym{rsrq}{RSRQ}{Reference Signal Received Quality}
\newacronym{rssi}{RSSI}{Received Signal Strength Indicator}
\newacronym{crs}{CRS}{Cell Reference Signal}
\newacronym{v2v}{V2V}{Vehicle-to-Vehicle}
\newacronym{v2i}{V2I}{Vehicle-to-Infrastructure}
\newacronym{v2x}{V2X}{Vehicle-to-Everything}
\newacronym{vn}{VN}{Vehicular Node}
\newacronym{dsrc}{DSRC}{Dedicated Short Range Communication}
\newacronym{ci}{CI}{context information}
\newacronym{voi}{VoI}{value of information}
\newacronym{gps}{GPS}{Global Positioning System}
\newacronym{qos}{QoS}{Quality of Service}
\newacronym{ml}{ML}{Machine Learning}
\newacronym{ahp}{AHP}{Analytic Hierarchy Process}
\newacronym{lidar}{LIDAR}{Light Detection and Ranging}
\usepackage{cite}
\usepackage{bbold}
\usepackage{bbm}

\def \eqA{
\begin{equation}
\label{eq:A}
\mathcal{A}= 
	\begin{cases} 6.9+20\log_{10}\big[ \sqrt{(v-0.1)^2+1}+v-0.1\big], & \mbox{if }v > -0.7 \\ 
	0, & \mbox{otherwise}
	\end{cases}
\end{equation}
}

\def \pldsrc{
\begin{align}
	PL(d)=&\mathcal{A}\mathbbm{1}_{N} +
	PL(d_0)-
	10\gamma_1\log_{10}\left\lfloor\tfrac{d}{d_0}\right\rceil_1^{\frac{d_c}{d_0}}-\\&-10\gamma_2\log_{10}\left\lfloor\tfrac{d}{d_c}\right\rceil_1^\infty+\chi_{\sigma},\notag
		\label{eq:pl_dsrc}
\end{align}
}

\usepackage{siunitx}
\usetikzlibrary{fadings}

\tikzfading[name=middle,
            top color=transparent!100,
            bottom color=transparent!100,
            middle color=transparent!20]

\usetikzlibrary{arrows,automata,calc,shapes, positioning,shadows,shadows.blur,shapes.geometric}

\makeglossaries

\begin{document}


\title{On the Feasibility of Integrating mmWave and \\ IEEE 802.11p for V2V Communications}

\author{\IEEEauthorblockN{ Marco Giordani$^{\circ }$, Andrea Zanella$^{\circ }$, Takamasa Higuchi$^{\dagger }$, Onur Altintas$^{\dagger}$, Michele Zorzi$^{\circ }$}
\IEEEauthorblockA{\\
$^{\circ }$Consorzio Futuro in Ricerca (CFR) and  University of Padova, Italy, 
 Email:{\{giordani,zanella,zorzi\}@dei.unipd.it, }\\
$^{\dagger}$TOYOTA InfoTechnology Center, Mountain View, CA, USA,
Email:{\{ta-higuchi,onur\}@us.toyota-itc.com }}}

\maketitle

\begin{abstract}
Recently, the millimeter wave (mmWave) band has been investigated as a means to support the foreseen extreme data rate demands of emerging automotive applications, which go beyond  the capabilities of existing  technologies for vehicular communications.
However, this potential is hindered by the severe isotropic path loss and the harsh propagation of high-frequency channels. 
Moreover, mmWave signals  are typically directional, to benefit from  beamforming gain, and require frequent re-alignment of the beams to maintain connectivity.
These limitations are particularly challenging when considering vehicle-to-vehicle (V2V) transmissions, because of the highly mobile nature of the vehicular scenarios, and pose new challenges for  proper vehicular communication design. 
 In this paper, we conduct simulations to compare the performance of IEEE 802.11p and the mmWave technology to support V2V networking, aiming at providing insights on how both technologies can complement each other to meet the requirements of future automotive services.
 The results show that  mmWave-based strategies support ultra-high transmission speeds, and IEEE 802.11p systems have the ability to guarantee   reliable and robust~communications.
\end{abstract}


\section{Introduction} 
\label{sec:introduction}
In recent years, \gls{v2x} communications have been investigated as a means to support  automotive services to improve the efficiency and safety of road transportation systems. 
These safety services often deal with small data messages with  very stringent requirements in terms of transmission reliability and latency~\cite{bookV2X}.
However, future connected and automated vehicles will encompass a wide range of sensors, including video cameras, GPS, LIDARs and radars, that will generate massive amounts of data (e.g., in the order of terabytes per driving hour~\cite{magazine2016_Heath}) that may exceed the capacity of existing \gls{v2x} communication technologies.

A possible response to this growing demand for ultra-high transmission speeds can be found in the next-generation radio interfaces, globally standardized by the \gls{3gpp} as \gls{nr}~\cite{3GPP_38802}, that include in particular the \gls{mmwave} bands.\footnote{Although strictly speaking \gls{mmwave} bands include frequencies between 30 and 300 GHz, industry has loosely defined it to include any frequency above 10 GHz.} 
Besides the extremely large bandwidths available at such frequencies, the small size of the antennas makes it possible to build complex antenna arrays and obtain high gains by \gls{bf}, thus further increasing the transmission rates.
The \gls{mmwave} band therefore represents  a new opportunity for future vehicular communications, in combination with  other wireless systems.
However, this potential can be jeopardized by the  challenging propagation characteristics  of high-frequency channels, as signals do not penetrate most solid materials and are subject to high signal attenuation and reflection.
In addition, mmWave links are typically directional to benefit from  beamforming gain and, especially when considering highly dense or highly~mobile vehicular scenarios, require precise alignment of the transmitter and receiver beams to maintain connectivity~\cite{MOCAST_2017,giordani2018coverage}.

\begin{table*}[!t]
\centering
\caption{Description of the characteristics of radio interfaces currently being considered for V2V communications. }
\label{tab:rats}
\renewcommand{\arraystretch}{1}
\begin{tabular}{p{2cm}p{6.3cm}p{6.3cm}}
\toprule
RT & Pros & Cons \\
\midrule
IEEE 802.11p \cite{li2010overview} &
\begin{tabular}[t]{@{\textbullet~}p{5.8cm}@{}}
 No need for network infrastructure\\
 Low latency\\
 Fully distributed and uncoordinated access
\end{tabular} &
\begin{tabular}[t]{@{\textbullet~}p{5.8cm}@{}}
 \emph{Hidden node} problem\\
 Limited transmission data rate	
\end{tabular} \\
\midrule
mmWaves \cite{rappaport2014millimeter} &
\begin{tabular}[t]{@{\textbullet~}p{5.8cm}@{}}
Very large bandwidth \\
 Beamforming gain\\
 Spatial isolation\\
Potential of physical-layer security/privacy
\end{tabular} &
\begin{tabular}[t]{@{\textbullet~}p{5.8cm}@{}}
 Very large path loss\\
 Signals do not penetrate through solid material\\
 Need to set up aligned transmissions\\
 Significant shadowing, reflection and scattering
\end{tabular} \\
\bottomrule
\end{tabular}
\end{table*}



In this regard, while the research on \gls{v2i} systems operating at \glspl{mmwave} has recently been quite widespread (e.g.,~\cite{beamDesignV2I_Heath,locationAided_Garcia,radarV2I_Heath,magazine2016_Heath,performance2018giordani}),  the literature on  \gls{v2v} networking (e.g.,~\cite{perfecto2017millimeter,hu2018vehicular}) is still very scarce.
In \cite{perfecto2017millimeter} the authors shed light on the operational limits of
\gls{mmwave} bands as a viable \gls{rt} for future high-rate \gls{v2v} transmissions while, in \cite{hu2018vehicular}, a multi-access edge computing framework  integrating licensed sub-6 GHz band  and \glspl{mmwave}  for inter-vehicle information distribution is proposed.
Some other papers have provided analytical characterizations of the \gls{mmwave} channel in \gls{v2v} scenarios. 
For example, in \cite{Tassi17_Highway}, the authors theoretically modeled the downlink performance of a mmWave network deployed along a highway section, while in \cite{sanchez2017millimeter} an experimental characterization of the 38 GHz and 60 GHz radio channel is presented.
Motivated by these considerations, in this paper we conduct extensive simulations to provide the first numerical evaluation of the practical feasibility of designing \gls{mmwave}-based strategies to target the requirements of future V2V  services, and  compare their performance to that of  IEEE 802.11p systems.
The results show that the support of  high-frequency bands has the potential to provide extremely high data rates, while IEEE 802.11p enables reliable and robust communications with lower bit-rates. 
We conclude that the orchestration of different radios makes it possible to complement the limitations of each type of network and therefore represents  a viable approach to improve the robustness and stability of  \gls{v2v} connectivity while establishing high-capacity~channels.

The remainder of this paper is organized as follows. 
In Sec.~\ref{sec:v2v_communications_dsrc_vs_mmwaves} we overview the characteristics of the IEEE 802.11p standard and the \gls{mmwave} technology in relation with  target \gls{v2v} application requirements.
The system model is described in Sec.~\ref{sec:system_model}, while in Sec.~\ref{sec:comparative_analysis_and_results} we present our main findings and simulation results.
Finally, conclusions and suggestions for future work are provided in~Sec.~\ref{sec:conclusions_and_future_work}.


\section{Radio  Technologies to Enable \\ \gls{v2v} Networking} 
\label{sec:v2v_communications_dsrc_vs_mmwaves}

\gls{v2v} communications are designed to exchange basic information among the vehicles to enable advanced automotive services, the main ones concerning the enhancement of road safety and the reduction of the traffic impact on the environment.
The  \gls{3gpp} has recently categorized different performance requirements for next-generation vehicular systems supporting enhanced \gls{v2v} applications~\cite{3GPP_22186}, as follows. 

\emph{Vehicle platooning} includes services that make it possible for a group of vehicles that follow the same trajectory to travel in close proximity to one another, nose-to-tail, at highway speeds. In addition to the strict latency requirement, the connection reliability and stability of these applications are also very critical.

\emph{Advanced driving} 
enables semi- or fully-automated driving and allows vehicles to coordinate their trajectories and maneuvers to guarantee safer traveling, collision avoidance, and improved traffic efficiency. While the size of the exchanged safety messages is typically small (up to a few thousands of bytes), latency must be very small to ensure prompt reactions to unpredictable events. 

The \emph{extended-sensor} functionality enables the direct dissemination of raw or processed data, gathered through local sensors, among vehicles, which can enhance the perception of the surrounding environment beyond what their own instrumentation can detect.
These services usually require  high-throughput connections (in the order of hundreds of megabits per second), due to the detailed nature of the shared contents, while some latency can be tolerated (depending on the degree of automation).

It is therefore clear that  future  \gls{v2v} services will have increasingly stringent demands in terms of data rate, reliability, latency, and connectivity.
In this section, we thus overview the features of candidate \glspl{rt} currently being considered for inter-vehicle communications, namely the IEEE 802.11p standard and the \gls{mmwave} technology,\footnote{Based on the \gls{3gpp} Release 14 specifications, the cellular-V2X~{(C-V2X)} technology, which relies on the PC5 interface specified for device-to-device operations, offers \gls{v2v} services using cellular technologies as a basis \cite{3GPP_23785}. However, a comparison between C-V2X and the IEEE 802.11p and the  \gls{mmwave} paradigms is beyond the scope of this paper.} and discuss their performance characteristics in relation with  target  application~requirements.
Table~\ref{tab:rats} provides a short summary of the~argumentation.

\subsection{IEEE 802.11p Communications} 
\label{ssub:dsrc_for_v2v_communications}
The IEEE 802.11p standard supports the \gls{phy} and \gls{mac} layers of the \gls{dsrc} transmission service and offers \gls{v2v} data exchange at a nominal rate  from 6 to 27 Mbps within a range of a few hundreds of meters~\cite{li2010overview}.
In the US, \gls{dsrc} can count on a total spectrum of 75 MHz in the 5.9 GHz frequency band, divided into seven 10-MHz channels, with 5 MHz of guard band at the lower end of the spectrum.
Japan has been already deployed IEEE 802.11p systems using 10 MHz of spectrum  in the 760 MHz band~\cite{toyota2015toyotabringing}.
This standard embeds certain desirable features for \gls{v2v} communications.
Endpoints can operate without a network infrastructure, removing the need for prior exchange of control information and thus bringing a significant advantage in terms of latency with respect to  regular Wi-Fi or legacy cellular operations. Moreover, IEEE 802.11p implements the carrier sensing multiple access with collision avoidance (CSMA/CA) mechanism at the \gls{mac} layer, thereby guaranteeing a fully distributed and uncoordinated access to the wireless channel, with no need for a resource allocation~procedure.

 Nevertheless, the IEEE 802.11p  standard presents some  inherent limitations.
 First, the throughput and delay performance degrades  as the network load increases, even though there are ways of mitigating congestion by adjusting the message rate in the application layer~\cite{bansal2013limeric}.
 Second, the channel access mechanism is prone to the \emph{hidden node} problem, which may result in packet collisions.
 Third, the limited bandwidth of IEEE 802.11p systems results in limited data rates which may not satisfy the requirements of some categories of future \gls{v2v} applications, e.g., extended sensors.

In conclusion, on the one hand,  IEEE 802.11p systems present desirable features in terms of ubiquitous connection availability and communication stability.
On the other hand, 
some next-generation \gls{v2v} use cases may require high-throughput transmissions beyond the capacity of existing DSRC~systems.


\subsection{Millimeter Wave Communications} 
\label{ssub:mmwaves_for_v2v_communications}
The \gls{mmwave} band between 10 GHz and 300 GHz has been considered as an enabler of the \gls{5g} performance requirements in micro and picocellular networks \cite{RanRapE:14}.
As mentioned, these frequencies offer much more bandwidth than traditional networks operating in the congested bands below 6 GHz, and some preliminary capacity estimations have demonstrated that systems operating at \glspl{mmwave} can offer orders of magnitude higher bitrates than legacy vehicular networks \cite{kato2001its}.
Moreover, the millimeter wavelengths  make it practical to build very large antenna arrays (e.g., with 32 or more elements)
to provide spatial isolation, reduce interference, and support multiplexing.

Although this new band has gathered great interest for automotive applications, there are many concerns about its transmission characteristics \cite{MOCAST_2017}.
First, with wavelengths in the order of millimeters, isotropic transmissions incur severe path loss and result in limited communication range. 
Second,
 unlike signals at sub-6 GHz frequencies, \gls{mmwave} signals do not penetrate most solid materials, and movements of obstacles and reflectors may cause the channel to rapidly appear and disappear. 
Third, the significant Doppler spread experienced at high frequencies may impair the feedback over a  broadcast~channel (e.g., during synchronization or random~access).

To overcome these limitations, contrary to legacy \gls{v2v} schemes, \gls{mmwave} communications are typically directional to benefit from  beamforming gain, and  support mechanisms by which the vehicles can quickly determine  appropriate directions of transmission.
This requires precise and continuous alignment of the transmitter and receiver beams to maintain connectivity, an operation that may increase the system overhead and lead to throughput degradation. 
In this regard, the definition of directional \gls{v2v} strategies can be favored by the dissemination of in-vehicle sensors information, e.g., vehicle position information obtained from GPS measurements may help geometrically select the optimal beam to interconnect the endpoints at any given time and alleviate the burden of the beam alignment operations.
However, the communication performance can be rather deteriorated if the data is  inaccurate, outdated and/or unreliable.
Moreover, albeit the implementation of a digital beamforming architecture allows the processing of multiple simultaneous and orthogonal beams in the digital domain and  offers considerably faster alignment operations, it requires a separate \gls{rf} chain for each antenna element and therefore suffers from significant energy consumption and requires expensive hardware~\cite{Waqas_EW2016}.

In conclusion, the design of  \gls{v2v} systems~operating at \glspl{mmwave} requires the  implementation of mechanisms able to cope with the inherent instability   of the high-frequency channels, a research challenge that is still largely~unexplored.

\section{System Model} 
\label{sec:system_model}
In this section we describe the system model we considered to evaluate the communication performance of a vehicular system implementing \gls{v2v} transmissions.
The path loss characterization is presented in Sec.~\ref{ssub:nlos} and Sec.~\ref{sub:ppath_loss_model}, while in Sec.~\ref{ssub:simulation_parameters} we introduce our main simulation parameters.


\subsection{Non Line of Sight (NLOS) Probability} 
\label{ssub:nlos}

In \gls{v2v} systems, due to the relatively low elevation of the vehicle antennas, we reasonably expect that other vehicles will act as obstacles to the propagation of the signals. It is thus imperative to distinguish between \gls{los} and \gls{nlos} path~loss~characterizations.\footnote{In order to better investigate the effect of dynamic blockages (i.e., vehicles) on the signal propagation, in our simulations we have not considered static obstacles, e.g., buildings or vegetation. A complete characterization of the NLOS path loss probability is however of great interest and is left for future~work.}

In this regard, in \cite{boban2016modeling} the authors represent the \emph{blockage probability} $P_{\rm b}(d)$, i.e., the probability of one or more vehicles potentially obstructing the \gls{los} between the transmitter and the receiver,  as a non-decreasing function of the inter-vehicle distance $d$ (the longer the link, the more likely to intersect with one or more blockages), i.e.,
\begin{equation}
	P_{\rm b}(d) = 1-\min(1,\max(0,ad^2+bd+c)),
\end{equation}
where the parameters $a,b,c$ are derived from  geometry-based deterministic simulations and depend on the vehicle density of the scenario.
Nonetheless, from an electromagnetic  perspective, the presence of intermediate vehicles potentially obstructing the visual line of sight between the endpoints does not necessarily imply the \gls{nlos} condition. 
It is also required that the \emph{Fresnel ellipsoid} is not free of obstructions.
Therefore, according to the analysis developed in \cite{boban2011impact}, the probability $P_{\rm NLOS}(d)$ of \gls{nlos} condition for a link that spans a distance $d$ is given by the blockage probability multiplied by the probability that at least one vehicle is within the ellipsoid corresponding to 60\% of the radius of the Fresnel zone, i.e., 
\begin{equation}
	P_{\rm NLOS}(d) = P_{\rm b}(d) \cdot 	Q\left( \frac{h-\mu_h}{\sigma_h} \right),
\end{equation}
where the Q-function $Q(\cdot)$ represents the  tail distribution of  the normal distribution  function, and $\mu_h$ and $\sigma_h$ are the mean and the standard deviation of the obstacle height, respectively. Finally, $h$ denotes the effective height of the straight line connecting the transmitter and the receiver at the obstacle  and is given by
\begin{equation}
	h = (h_i-h_j)\frac{d_{\rm obs}}{d} + h_i - 0.6r_f+\ell_a,
\end{equation}
where $h_i$ and $h_j$ are the heights of the transmitting and the receiving vehicles, respectively, $d_{\rm obs}$,  which is uniformly distributed in [0, $d$], is the distance between the transmitter and the obstacle, $\ell_a$ is the physical length of the antenna and  $r_f$ is the radius of the first Fresnel ellipsoid, which is given~by
\begin{equation}
	r_f = \sqrt{\frac{\lambda d_{\rm obs}(d-d_{\rm obs})}{d}},
	\label{eq:r_f}
\end{equation}
with $\lambda$ denoting the wavelength.

\subsection{Path Loss Model} 
\label{sub:ppath_loss_model}

As soon as the different communication states (i.e., \gls{los} and \gls{nlos}) have been identified, the path loss  follows a \emph{dual-slope} piecewise-linear model, which is deemed suitable to represent the real propagation in a vehicular environment.
 We distinguish between IEEE 802.11p and \gls{mmwave} systems.

\textbf{IEEE 802.11p Model.}
First, we define the term $\left\lfloor x\right\rceil_a^b$~as
\begin{equation}
	\left\lfloor x\right\rceil_a^b\triangleq	
	\begin{cases} x, &\mbox{if } a\leq x\leq b \\ 
	a, &\mbox{if }  x< a \\ 
	b, &\mbox{if }  x> b \\ 
	\end{cases}
\end{equation}
Based on the above notation,  the path loss for IEEE 802.11p systems follows the model in~\cite{cheng2007mobile} and is expressed~as 
\pldsrc
where $\gamma_1$ and $\gamma_2$ are the path loss exponents, $\chi_{\sigma}$ represents the standard deviation of the shadowing, and $PL(d_0)$ is the free space path loss at the reference distance $d_0=1$ m. 
The critical parameter  $d_c$ is the Fresnel distance and is  calculated~as
\begin{equation}
	d_c = \frac{4h_{i}h_{j}}{\lambda}
\end{equation}
while the binary random variable $\mathbbm{1}_{N}$ is equal to 1 with probability $P_{\rm NLOS}(d)$, i.e., in case of \gls{nlos} propagation. In such a situation, the path loss is increased by a factor $\mathcal{A}$ according to a \emph{knife-edge} attenuation model\footnote{For the tractability of the simulation, we assume that the attenuation factor $\mathcal{A}$ follows a \emph{single knife-edge} model, which considers one single vehicle obstructing the \gls{los}. The extension of the single knife-edge obstacle case to a multiple knife-edge is not immediate and is left as future work.
}~\cite{boban2011impact},~i.e., 
\medmuskip=0mu
\thickmuskip=0mu
\eqA
\medmuskip=6mu
\thickmuskip=6mu
where $v=\sqrt{2}H/r_f$ and $H$ is the difference between the height of the obstacle and that of the line connecting the transmitting and the receiving vehicles.

\textbf{Millimeter-Wave Model.}
Available measurements at \glspl{mmwave} in the \gls{v2v} context are very limited and realistic scenarios are indeed hard to simulate. In fact, the increased reflectivity and scattering from common objects and the poor diffraction and penetration capabilities of \glspl{mmwave} are the main factors preventing the reuse of the existing sub-6 GHz path loss models for high-frequency scenarios~\cite{MOCAST_2017}.
Nevertheless, the authors in \cite{yamamoto2008path} have conducted some measurements at 60 GHz to characterize the \gls{mmwave} propagation between two cars communicating in \gls{los} or \gls{nlos} situations. The path loss is calculated according to a  dual-slope  model, i.e., 
\begin{equation}
	PL(d) = \xi\cdot 10\log_{10}d+\eta+15\cdot d/1000,
	\label{eq:PL_mmW}
\end{equation}
where the rightmost term denotes the atmospheric attenuation at 60 GHz, that is 15 dB/km, and the parameters $\xi$ and $\eta$ are given in~\cite[Table VI]{yamamoto2008path}. 
In the case of \gls{los} (with probability $P_{\rm LOS}(d) = 1-P_{\rm NLOS}(d)$), $\xi=1.77$ and $\eta=70$, while in the  case of \gls{nlos}, as for our previous assumptions,  we consider one single intermediate vehicle located midway between the transmitting and receiving vehicles, and we set $\xi=1.71$ and~${\eta=78.6}$.

\renewcommand{\arraystretch}{0.7}
\begin{table}[!t]
\small
\centering
\caption{Main simulation parameters.}
\begin{tabularx}{0.96\columnwidth}{ @{\extracolsep{\fill}} lll}
\toprule
Parameter & Value & Description \\ \midrule
$w$ & $3.5$ m & Lane width \\
$N_l$ & $4$  & Number of lanes \\
$P_{\rm TX}$ & $19.5$ dBm & Transmission power    \\
$h_i=h_j$ & $142$ cm & Height of TX and RX cars\\
$\mu_h$ & $150$ cm & Mean of intermediate car's height\\
$\sigma_h$ & $8.4$ cm & Stddev. of intermediate car's height\\
$d$ & \{2,\dots,500\} m & Inter-vehicle distance\\
\midrule
 $W_{\rm mmW}$ & $400$ MHz & mmWave total bandwidth \\
$f_{\rm c, mmW}$ & $60$ GHz &  mmWave carrier frequency \\
$N$ & $\{4,64\}$  & Antenna array size  \\
\midrule
 $W_{\rm DSRC}$ & $75$ MHz & IEEE 802.11p total bandwidth \\
$f_{\rm c, DSRC}$ & $5.9$ GHz &  IEEE 802.11p carrier frequency \\
$\ell_a$ & $10$ cm & IEEE 802.11p antenna length \\
\midrule
\multicolumn{3}{c}{$\gamma_1,\gamma_2,\xi_1,\chi_2$ (IEEE 802.11p path loss parameters) ~$\sim$\cite{cheng2007mobile} }  \\
\multicolumn{3}{c}{$\xi,\eta$ (mmWave path loss parameters) ~$\sim$\cite{yamamoto2008path} }\\
\multicolumn{3}{c}{$a,b,c$ (blockage probability parameters) ~$\sim$\cite{boban2016modeling} }  \\
\bottomrule
\end{tabularx}
\label{tab:params}
\end{table}



\subsection{Simulation Parameters} 
\label{ssub:simulation_parameters}

The simulation parameters are based on realistic system design considerations and are summarized in Table~\ref{tab:params}.
Vehicles are deployed over a section of a highway, which is composed of $N_l = 4$ infinitely long parallel lanes of width $w=3.5$~m, so that  the total road width is $2W = N_l\cdot w = 14$~m. 
We consider a static scenario, therefore the impact of the vehicles' speed is not investigated at this early stage.
Following the characterization proposed in the prior literature, the transmitting and  receiving vehicles are modeled as sedan cars of height $h_i = h_j = 142$ cm \cite{yamamoto2008path} while, in the case of \gls{nlos}, the intermediate vehicle height is modeled as a normal random variable with mean $\mu_h = 150$ cm and standard deviation $\sigma_h = 8.4$ cm~\cite{boban2011impact}.
 
 IEEE 802.11p systems operate in the legacy band, i.e., at 5.9~GHz, with  a total bandwidth of 75 MHz. Antennas are supposed to be located on top of vehicles, in the middle of the roof (which was experimentally shown to be the overall optimum placement when considering omnidirectional transmissions~\cite{boban2011impact}), and to be of length $\ell_a = 10$ cm.

 For \gls{mmwave} links, the central frequency is set to  60~GHz while the total bandwidth is set to  400~MHz, as specified in \cite{3GPP_38802}. 
In order to establish directional transmissions, vehicles are equipped with  Uniform Planar Arrays (UPAs) of $N$ elements, allowing to steer beams consisting of a main lobe of predefined width (which depends on $N$) and a side lobe that covers the remainder of the antenna radiation pattern.\footnote{It is quite reasonable to assume that four equal and independent antenna arrays are arranged on either side of each car and that the cars are positioned onto the same elevation plane, so that each array is responsible for covering a $\Delta_\theta = 180^\circ$ azimuth space.}
For completeness, in our study we also consider omnidirectional \gls{mmwave} transmissions, i.e., $N=1$.
For the beam alignment, we assume that measurement reports are periodically exchanged (i.e., at the beginning of every slot of duration $T$) among the vehicles so that they can periodically identify the optimal directions for their respective beams~\cite{giordani2018coverage}.  Such configuration is kept fixed for the whole slot, during which nodes may lose the alignment due to mobility. 
 In case the connectivity is lost during a slot, it can only be recovered at the beginning of the subsequent slot, when the re-alignment procedure is performed again~\cite{giordani2018coverage}.
In this respect,  geographical position data can be used to geometrically point the  beam towards  the intended receiver at any given time: the inaccuracy of such  data is modeled according to a Gamma distribution with parameters $\alpha = 3.14733$ and $\beta = 0.462432$~\cite{driver2007long}.

Our results are derived through a Monte Carlo approach where $N_{\rm sim}=100000$ independent simulations are repeated to get different statistical quantities of interest.
In particular, we analyze the received signal strength between the transmitting and the receiving nodes for different values of the inter-vehicle distance $d$, with $d$ varying from 2 m to 500 m.
We recall that, with our channel model, the presence of the line of sight is probabilistically determined, thereby making the Monte Carlo approach a reasonable solution.


\section{Comparative Results} 
\label{sec:comparative_analysis_and_results}


In this section we provide some numerical results to compare the performance of  IEEE 802.11p and mmWave-based \gls{v2v} communications, which will be assessed in terms of achievable data rate, outage probability and~stability.\\

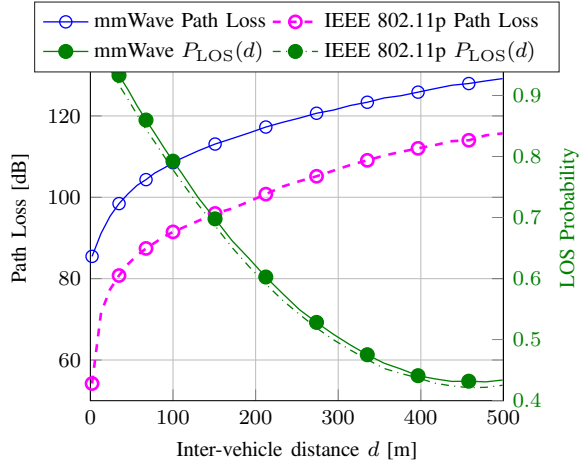
\begin{figure}[t!]
     \centering
     		\setlength{\belowcaptionskip}{-0.5cm}
      		\setlength{\belowcaptionskip}{0cm}
	\setlength{\belowcaptionskip}{0cm}
	\setlength\fwidth{0.62\columnwidth}
	\setlength\fheight{0.55\columnwidth}
%
%
\definecolor{mycolor1}{rgb}{1.00000,0.00000,1.00000}%
\pgfplotsset{
tick label style={font=\footnotesize},
label style={font=\footnotesize},
legend  style={font=\footnotesize}
}
\begin{tikzpicture}

\begin{axis}[%
width=\fwidth,
height=\fheight,
at={(0\fwidth,0\fheight)},
scale only axis,
xmin=0,
xmax=500,
xlabel style={font=\color{white!15!black}},
xlabel={Inter-vehicle distance $d$ [m]},
ymin=50,
ymax=140,
yminorticks=true,
ylabel style={font=\color{white!15!black}},
ylabel={Path Loss [dB]},
axis background/.style={fill=white},
xmajorgrids,
ymajorgrids,
yminorgrids,
mark repeat={3},
label style={font=\footnotesize},
legend columns={2},
legend style={at={(0.5,1.09)},legend cell align=left, align=left, anchor = north, draw=white!15!black},
]

\addplot [color=mycolor1, dashed, line width=1pt, mark size=2.3pt, mark=o, mark options={solid, mycolor1}, forget plot]
  table[row sep=crcr]{%
2	54.213356311383\\
12.8888888888889	71.6135110611529\\
23.7777777777778	77.3398198902921\\
34.6666666666667	80.7657711310336\\
45.5555555555556	83.5910316634053\\
56.4444444444444	85.687571393821\\
67.3333333333333	87.4290334580636\\
78.2222222222222	88.8630789499178\\
89.1111111111111	90.3273517358384\\
100	91.4906843286554\\
110	92.3562893761783\\
130.473684210526	94.19445085734\\
150.947368421053	96.0306383424592\\
171.421052631579	97.2109513403021\\
191.894736842105	98.9533812623036\\
212.368421052632	100.776682437605\\
232.842105263158	102.651435745408\\
253.315789473684	103.96979651643\\
273.789473684211	105.166276397627\\
294.263157894737	106.552905554512\\
314.736842105263	107.941806338853\\
335.210526315789	109.085730508785\\
355.684210526316	110.26155787136\\
376.157894736842	111.174955429877\\
396.631578947368	112.02921611785\\
417.105263157895	112.831656991123\\
437.578947368421	113.801187351878\\
458.052631578947	114.011565464579\\
478.526315789474	115.22066857503\\
499	115.70466766166\\
};

\addplot [color=blue, line width=0.5pt, mark size=2.3pt, mark=o, mark options={solid, blue}, forget plot]
  table[row sep=crcr]{%
2	85.4741184184169\\
12.8888888888889	91.1249994499796\\
23.7777777777778	95.3367345560049\\
34.6666666666667	98.4468152710629\\
45.5555555555556	100.844199871733\\
56.4444444444444	102.926813923495\\
67.3333333333333	104.339137467032\\
78.2222222222222	106.004252387252\\
89.1111111111111	107.185768558068\\
100	108.491122685192\\
110	109.541935772669\\
130.473684210526	111.416898030352\\
150.947368421053	113.086179531351\\
171.421052631579	114.431269596113\\
191.894736842105	115.811698923189\\
212.368421052632	117.269567234923\\
232.842105263158	118.438840244444\\
253.315789473684	119.484017155975\\
273.789473684211	120.657542881595\\
294.263157894737	121.445540164898\\
314.736842105263	122.573590595548\\
335.210526315789	123.36353024766\\
355.684210526316	124.4221646288\\
376.157894736842	124.974791417962\\
396.631578947368	125.843725728558\\
417.105263157895	126.713023900966\\
437.578947368421	127.54245176714\\
458.052631578947	127.962641121181\\
478.526315789474	128.606266513637\\
499	129.173239571791\\
};

\end{axis}

\begin{axis}[%
width=\fwidth,
height=\fheight,
at={(0\fwidth,0\fheight)},
scale only axis,
axis y line*=right,
xmin=0,
xmax=500,
  axis x line=none,
  xlabel style={font=\color{white!15!black}},
   extra y ticks={0.5,0.7, 0.9},
  every outer y axis line/.append style={font=\footnotesize \color black!50!green},
every y tick label/.append style={font=\footnotesize \color{black!50!green}},
every y tick/.append style={font=\footnotesize \color black!50!green},
xlabel={Inter-vehicle distance $d$ [m]},
ymin=0.4,
ymax=1,
yminorticks=true,
mark repeat={3},
ylabel={LOS Probability},
label style={font=\footnotesize},
ylabel style={font=\footnotesize \color{black!50!green}},
legend columns={2},
legend style={at={(0.5,1.09)},legend cell align=left, align=left, anchor = north, draw=white!15!black},
]
\addplot [color=black!50!green, dash dot, line width=0.5pt, forget plot]
  table[row sep=crcr]{%
2	0.99628\\
12.8888888888889	0.97\\
23.7777777777778	0.94348\\
34.6666666666667	0.91735\\
45.5555555555556	0.89377\\
56.4444444444444	0.86892\\
67.3333333333333	0.84426\\
78.2222222222222	0.82311\\
89.1111111111111	0.80071\\
100	0.77836\\
110	0.75918\\
130.473684210526	0.72118\\
150.947368421053	0.68666\\
171.421052631579	0.65101\\
191.894736842105	0.61991\\
212.368421052632	0.59095\\
232.842105263158	0.56425\\
253.315789473684	0.54109\\
273.789473684211	0.51906\\
294.263157894737	0.49966\\
314.736842105263	0.48291\\
335.210526315789	0.46724\\
355.684210526316	0.45258\\
376.157894736842	0.44436\\
396.631578947368	0.43706\\
417.105263157895	0.42849\\
437.578947368421	0.42396\\
458.052631578947	0.42126\\
478.526315789474	0.42169\\
499	0.42512\\
};

\addplot [color=black!50!green, line width=0.5pt,mark size=2.6pt, mark=*, mark options={solid, black!50!green},forget plot]
  table[row sep=crcr]{%
2 0.99767\\
12.8888888888889  0.97825\\
23.7777777777778  0.95559\\
34.6666666666667  0.93293\\
45.5555555555556  0.90957\\
56.4444444444444  0.88342\\
67.3333333333333  0.85974\\
78.2222222222222  0.83753\\
89.1111111111111  0.81438\\
100 0.79237\\
110 0.77391\\
130.473684210526  0.73274\\
150.947368421053  0.69782\\
171.421052631579  0.66221\\
191.894736842105  0.6331\\
212.368421052632  0.60251\\
232.842105263158  0.57519\\
253.315789473684  0.54878\\
273.789473684211  0.52811\\
294.263157894737  0.50862\\
314.736842105263  0.49113\\
335.210526315789  0.47514\\
355.684210526316  0.4612\\
376.157894736842  0.45163\\
396.631578947368  0.44079\\
417.105263157895  0.43559\\
437.578947368421  0.4314\\
458.052631578947  0.43177\\
478.526315789474  0.43067\\
499 0.43368\\
};

\addplot [color=blue, line width=0.5pt, mark size=2.3pt, mark=o, mark options={solid, blue}]
  table[row sep=crcr]{%
0 0\\
  };
  \addlegendentry{mmWave Path Loss}

  \addplot [color=mycolor1, dashed, line width=1pt, mark size=2.3pt, mark=o, mark options={solid, mycolor1}]
  table[row sep=crcr]{%
0 0\\
  };
  \addlegendentry{IEEE 802.11p Path Loss }

  \addplot [color=black!50!green, line width=0.5pt, mark size=2.6pt, mark=*, mark options={solid, black!50!green}]
  table[row sep=crcr]{%
0 0\\
  };
  \addlegendentry{mmWave $P_{\rm LOS}(d)$}

    \addplot [color=black!50!green, dash dot, line width=0.5pt, mark size=2.6pt, mark=*, mark options={solid, black!50!green}]
  table[row sep=crcr]{%
0 0\\
  };
  \addlegendentry{IEEE 802.11p $P_{\rm LOS}(d)$}

\end{axis}
\end{tikzpicture}%
    \caption{\footnotesize  Path loss (left axis) and LOS probability (right axis)  \emph{vs} $d$ for IEEE 802.11p  and \gls{mmwave} communications. }
    \label{fig:PL}
      \end{figure}

\textbf{Path Loss.} Fig.~\ref{fig:PL} shows the path loss (left axis) and the \gls{los} probability (right axis) as a function of the inter-vehicle distance $d$ and considering both IEEE 802.11p and \gls{mmwave} transmissions.
As expected, the LOS probability $P_{\rm LOS}(d)=1-P_{\rm NLOS}(d)$ is a non-increasing function of $d$ as the farther apart the vehicles, the more likely the event of obstruction of the line of sight.
We notice that $P_{\rm LOS}(d)$ is slightly lower for IEEE 802.11p systems since, when operating at lower frequencies, the Fresnel radius $r_f$ increases, as highlighted by Eq.~\eqref{eq:r_f}, thereby increasing the probability of vehicles obstructing the Fresnel ellipsoid.
Nonetheless,  the mmWave path loss is significantly higher than for IEEE 802.11p transmissions.
The reason is that, unlike  sub-6~GHz frequencies, \glspl{mmwave} have increased reflectivity, poor diffraction and penetration capabilities in \gls{nlos} situations, and therefore are affected by significant attenuation.
However,  the effect of such properties is small for short distances (i.e., up to a few tens of meters), which therefore represent a suitable range for \gls{mmwave}  links in vehicular~scenarios.\\

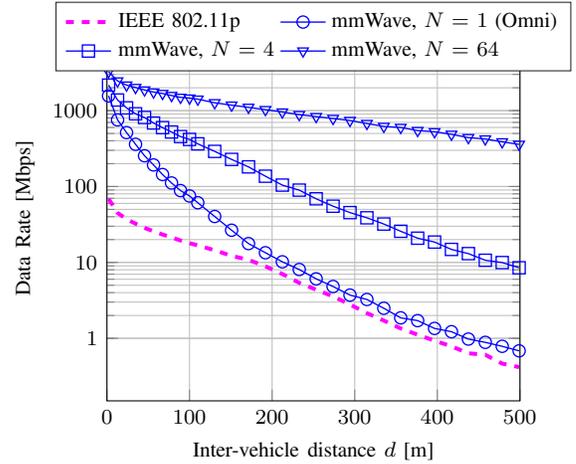
\begin{figure}[t!]
     \centering
     		\setlength{\belowcaptionskip}{-0.5cm}
      		\setlength{\belowcaptionskip}{0cm}
	\setlength{\belowcaptionskip}{0cm}
	\setlength\fwidth{0.62\columnwidth}
	\setlength\fheight{0.55\columnwidth}
%
%
\definecolor{mycolor1}{rgb}{1.00000,0.00000,1.00000}%
\pgfplotsset{
tick label style={font=\footnotesize},
label style={font=\footnotesize},
legend  style={font=\footnotesize}
}
\begin{tikzpicture}

\begin{axis}[%
width=\fwidth,
height=\fheight,
at={(0\fwidth,0\fheight)},
scale only axis,
xmin=0,
xmax=500,
xlabel style={font=\color{white!15!black}},
xlabel={Inter-vehicle distance $d$ [m]},
ymode=log,
ymin=0.150000,
ymax=10000.000000,
yminorticks=true,
ylabel style={font=\color{white!15!black}},
ylabel={Data Rate [Mbps]},
axis background/.style={fill=white},
ytick={1,  10,100,1000},
yticklabels={$1$,$10$,$100$,$1000$},
xmajorgrids,
ymajorgrids,
yminorgrids,
label style={font=\footnotesize},
legend columns={2},
legend style={legend cell align=left, align=left, at={(0.5,1.09)}, anchor = north, draw=white!15!black},
]

\addplot [color=mycolor1, dashed, line width=1.5pt]
  table[row sep=crcr]{%
2	68.8951136849663\\
12.8888888888889	44.9624407955596\\
23.7777777777778	37.0940778597223\\
34.6666666666667	32.3957147890859\\
45.5555555555556	28.5333873432793\\
56.4444444444444	25.6805303969036\\
67.3333333333333	23.3248678488392\\
78.2222222222222	21.3991070322261\\
89.1111111111111	19.4511157386394\\
100	17.9211477131254\\
110	16.7956746298674\\
130.473684210526	14.4534060887891\\
150.947368421053	12.2009198211585\\
171.421052631579	10.8145403799132\\
191.894736842105	8.88166215949394\\
212.368421052632	7.03972019525264\\
232.842105263158	5.37909324613082\\
253.315789473684	4.37001664752747\\
273.789473684211	3.57273832805425\\
294.263157894737	2.78844878787265\\
314.736842105263	2.14515716618001\\
335.210526315789	1.71252350698449\\
355.684210526316	1.34849589391601\\
376.157894736842	1.11503144750036\\
396.631578947368	0.930585733023448\\
417.105263157895	0.783410869522512\\
437.578947368421	0.634672580306993\\
458.052631578947	0.606132056125877\\
478.526315789474	0.464393059310292\\
499	0.417086772542872\\
};
\addlegendentry{IEEE 802.11p}

\addplot [color=blue, line width=0.5pt, mark size=2.3pt, mark=o, mark options={solid, blue}]
  table[row sep=crcr]{%
2	1563.37487286092\\
12.8888888888889	754.887971415623\\
23.7777777777778	513.543014313119\\
34.6666666666667	360.226489402578\\
45.5555555555556	254.228885948736\\
56.4444444444444	192.537226284189\\
67.3333333333333	143.952075172537\\
78.2222222222222	111.372381217623\\
89.1111111111111	88.4210021937\\
100	75.1111996594736\\
110	61.1893629391786\\
130.473684210526	40.2718422670711\\
150.947368421053	26.5579665733037\\
171.421052631579	17.7065410861336\\
191.894736842105	13.4107317264088\\
212.368421052632	10.1848237002384\\
232.842105263158	8.06284823082955\\
253.315789473684	6.07388372522479\\
273.789473684211	4.81398278181742\\
294.263157894737	3.71228955552049\\
314.736842105263	3.24835159353552\\
335.210526315789	2.489314740416\\
355.684210526316	1.8608790258252\\
376.157894736842	1.71093542073992\\
396.631578947368	1.35017940544996\\
417.105263157895	1.22211224823273\\
437.578947368421	0.977809076940254\\
458.052631578947	0.884846862264677\\
478.526315789474	0.78463033047265\\
499	0.684087987796091\\
};
\addlegendentry{mmWave, $N=1$ (Omni)}

\addplot [color=blue, line width=0.5pt, mark size=2.3pt, mark=square, mark options={solid, blue}]
  table[row sep=crcr]{%
2	2170.63988870718\\
12.8888888888889	1379.59213191539\\
23.7777777777778	1087.86078818874\\
34.6666666666667	908.206323032297\\
45.5555555555556	810.839966388021\\
56.4444444444444	686.609374609894\\
67.3333333333333	597.34614410398\\
78.2222222222222	522.994824755909\\
89.1111111111111	454.959504149712\\
100	417.325341873275\\
110	365.093321376275\\
130.473684210526	290.392879121214\\
150.947368421053	227.63719052566\\
171.421052631579	181.491982729458\\
191.894736842105	136.746420908166\\
212.368421052632	104.355732486887\\
232.842105263158	89.658572937075\\
253.315789473684	68.8965570182687\\
273.789473684211	55.0622401032572\\
294.263157894737	45.3104349212054\\
314.736842105263	38.7220356191785\\
335.210526315789	32.0430384532032\\
355.684210526316	25.7162914472441\\
376.157894736842	20.8568339347536\\
396.631578947368	18.5503950081707\\
417.105263157895	14.8308396148421\\
437.578947368421	13.0597903950241\\
458.052631578947	10.6782376177457\\
478.526315789474	9.95869296561723\\
499	8.50140639323648\\
};
\addlegendentry{mmWave, $N = 4$}

\addplot [color=blue, line width=0.5pt, mark size=2.3pt, mark=triangle, mark options={solid, rotate=180, blue}]
  table[row sep=crcr]{%
2	3260.91949769774\\
12.8888888888889	2458.41238538859\\
23.7777777777778	2170.48231818831\\
34.6666666666667	2016.27214936875\\
45.5555555555556	1870.22576157485\\
56.4444444444444	1755.55932306405\\
67.3333333333333	1681.76909634067\\
78.2222222222222	1581.09681918715\\
89.1111111111111	1496.00255431636\\
100	1462.07296230891\\
110	1415.41888368466\\
130.473684210526	1269.47636420803\\
150.947368421053	1180.00848448508\\
171.421052631579	1104.94777066663\\
191.894736842105	1029.93333566872\\
212.368421052632	958.384352832389\\
232.842105263158	885.103704847874\\
253.315789473684	833.917750362704\\
273.789473684211	786.77214644693\\
294.263157894737	741.192388441682\\
314.736842105263	677.06756139586\\
335.210526315789	618.74015815737\\
355.684210526316	599.558791500017\\
376.157894736842	546.084745171947\\
396.631578947368	526.549371548609\\
417.105263157895	487.53409722359\\
437.578947368421	440.803654123596\\
458.052631578947	423.154286749254\\
478.526315789474	387.585191215018\\
499	359.268499210684\\
};
\addlegendentry{mmWave, $N = 64$}

\end{axis}
\end{tikzpicture}%
    \caption{\footnotesize  Average achievable data rate \emph{vs} $d$ for IEEE 802.11p (dashed line) and \gls{mmwave} communications (solid lines), with different antenna array configurations (markers).}
    \label{fig:datarate}
      \end{figure}

\textbf{Data Rate.} In Fig.~\ref{fig:datarate}, we compare the average Shannon data rate of both the IEEE 802.11p and the \gls{mmwave} technologies, for different antenna configurations (including omnidirectional mmWave transmissions, i.e., $N=1$).\footnote{In this paper, the Shannon data rate is computed from the \gls{snr}, which is the average  received power divided by the noise power. Therefore, our results represent an \emph{upper bound} for the performance of the vehicular nodes, as we do not investigate the effect of  interference nor  make any  medium access control consideration.}
We observe that the very large bandwidth available to the \gls{mmwave} systems (5 times larger than in IEEE 802.11p) ensures much higher throughput than operating at legacy frequencies (up to two orders of magnitude more in short range).
This performance can be further magnified by configuring very directional transmissions.
In fact, there exists a strong correlation among  beamwidth, number of antenna elements and beamforming gain: the more antenna elements in the system, the narrower the beams, the more  directional the transmission, the higher the gain that can be achieved by beamforming.

It should also be noted that, even  implementing~omnidirectional strategies at  \glspl{mmwave}, the connection still guarantees acceptable average bitrate, provided that the endpoints are sufficiently close (to increase the \gls{los}~probability).\\



\begin{figure}[t!]
     \centering
     		\setlength{\belowcaptionskip}{-0.5cm}
      		\setlength{\belowcaptionskip}{0cm}
	\setlength{\belowcaptionskip}{0cm}
	\setlength\fwidth{0.62\columnwidth}
	\setlength\fheight{0.55\columnwidth}
%
%
\definecolor{mycolor1}{rgb}{1.00000,0.00000,1.00000}%
\pgfplotsset{
tick label style={font=\footnotesize},
label style={font=\footnotesize},
legend  style={font=\footnotesize}
}
\begin{tikzpicture}

\begin{axis}[%
width=\fwidth,
height=\fheight,
at={(0\fwidth,0\fheight)},
scale only axis,
xmin=0,
xmax=500,
ymode=log,
ymin=0.001,
ymax=2,
yminorticks=true,
axis background/.style={fill=white},
xmajorgrids,
ymajorgrids,
yminorgrids,
ylabel={Outage probability},
xlabel style={font=\color{white!15!black}},
xlabel={Inter-vehicle distance $d$ [m]},
label style={font=\footnotesize},
legend columns={2},
legend style={at={(0.5,1.12)},legend cell align=left, align=left, anchor = north, draw=white!15!black},
]

\addplot [color=mycolor1, dashed, line width=1.5pt]
  table[row sep=crcr]{%
2	0\\
12.8888888888889	0\\
23.7777777777778	0\\
34.6666666666667	0\\
45.5555555555556	0\\
56.4444444444444	0\\
67.3333333333333	0\\
78.2222222222222	4e-05\\
89.1111111111111	1e-05\\
100	3e-05\\
110	4e-05\\
130.473684210526	4e-05\\
150.947368421053	7e-05\\
171.421052631579	9e-05\\
191.894736842105	0.00024\\
212.368421052632	0.00047\\
232.842105263158	0.00103\\
253.315789473684	0.00232\\
273.789473684211	0.00458\\
294.263157894737	0.00935\\
314.736842105263	0.01619\\
335.210526315789	0.02593\\
355.684210526316	0.03973\\
376.157894736842	0.05908\\
396.631578947368	0.08186\\
417.105263157895	0.10899\\
437.578947368421	0.13975\\
458.052631578947	0.1731\\
478.526315789474	0.20836\\
499	0.246\\
};
\addlegendentry{IEEE 802.11p}

\addplot [color=blue, line width=0.5pt, mark size=2.5pt, mark=o, mark options={solid, blue}]
  table[row sep=crcr]{%
2	0.0002\\
12.8888888888889	0.0027\\
23.7777777777778	0.0111\\
34.6666666666667	0.0234\\
45.5555555555556	0.0486\\
56.4444444444444	0.0631\\
67.3333333333333	0.1021\\
78.2222222222222	0.1456\\
89.1111111111111	0.2028\\
100	0.2415\\
110	0.2891\\
130.473684210526	0.4144\\
150.947368421053	0.5238\\
171.421052631579	0.5975\\
191.894736842105	0.7146\\
212.368421052632	0.7889\\
232.842105263158	0.863\\
253.315789473684	0.9132\\
273.789473684211	0.956\\
294.263157894737	0.9799\\
314.736842105263	0.9925\\
335.210526315789	0.9976\\
355.684210526316	0.999\\
376.157894736842	1\\
396.631578947368	0.9999\\
417.105263157895	1\\
437.578947368421	1\\
458.052631578947	1\\
478.526315789474	1\\
499	1\\
};
\addlegendentry{mmWave, $N=1$ (Omni)}

\addplot [color=blue, line width=0.5pt, mark size=2.5pt, mark=square, mark options={solid, blue}]
  table[row sep=crcr]{%
2	0\\
12.8888888888889	0.0003\\
23.7777777777778	0.0013\\
34.6666666666667	0.0018\\
45.5555555555556	0.0038\\
56.4444444444444	0.0051\\
67.3333333333333	0.0089\\
78.2222222222222	0.0112\\
89.1111111111111	0.0175\\
100	0.0231\\
110	0.0321\\
130.473684210526	0.0436\\
150.947368421053	0.0615\\
171.421052631579	0.0882\\
191.894736842105	0.1164\\
212.368421052632	0.1522\\
232.842105263158	0.1974\\
253.315789473684	0.2482\\
273.789473684211	0.3103\\
294.263157894737	0.3905\\
314.736842105263	0.4406\\
335.210526315789	0.5013\\
355.684210526316	0.5728\\
376.157894736842	0.6267\\
396.631578947368	0.6716\\
417.105263157895	0.704\\
437.578947368421	0.744\\
458.052631578947	0.7763\\
478.526315789474	0.8105\\
499	0.831\\
};
\addlegendentry{mmWave, $N = 4$}

\addplot [color=blue, line width=0.5pt, mark size=2.5pt, mark=triangle, mark options={solid, rotate=180, blue}]
  table[row sep=crcr]{%
2	0\\
12.8888888888889	7e-05\\
23.7777777777778	0.00014\\
34.6666666666667	0.00015\\
45.5555555555556	0.0003\\
56.4444444444444	0.00036\\
67.3333333333333	0.0004\\
78.2222222222222	0.00063\\
89.1111111111111	0.00073\\
100	0.00089\\
110	0.00079\\
130.473684210526	0.00104\\
150.947368421053	0.00175\\
171.421052631579	0.00267\\
191.894736842105	0.00326\\
212.368421052632	0.0042\\
232.842105263158	0.00491\\
253.315789473684	0.00595\\
273.789473684211	0.00746\\
294.263157894737	0.00881\\
314.736842105263	0.01063\\
335.210526315789	0.01201\\
355.684210526316	0.01428\\
376.157894736842	0.01625\\
396.631578947368	0.01828\\
417.105263157895	0.02164\\
437.578947368421	0.02422\\
458.052631578947	0.02863\\
478.526315789474	0.03305\\
499	0.03606\\
};
\addlegendentry{mmWave, $N = 64$}

\end{axis}
\end{tikzpicture}%
    \caption{\footnotesize  Outage probability \emph{vs} $d$ for IEEE 802.11p (dashed line) and \gls{mmwave} communications (solid lines), with different antenna array configurations (markers). }
    \label{fig:outage}
      \end{figure}
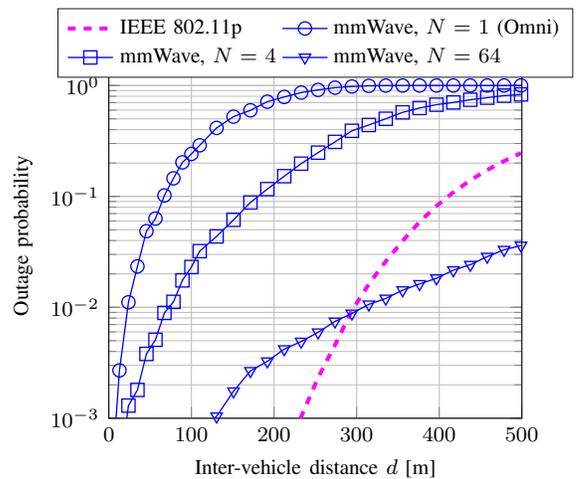

\textbf{Outage Probability.} 
In Fig.~\ref{fig:outage}, we evaluate the  outage probability of the investigated \gls{v2v} schemes,   i.e., the probability that the received signal strength  is below a predefined threshold, taken to be $-5$ dB in our simulations.
Low values of outage ensure more reliable  \gls{v2v} communications, a critical prerequisite for safety services requiring ubiquitous and continuous connectivity.

In general, we see that lower outage probability 
can be achieved when considering short-range communications and, in case of directional transmissions,  when using
large arrays. 
In the first case, the endpoints are progressively
closer, thus ensuring better signal quality and
stronger received power. 
In this region, the
channel conditions are sufficiently good to ensure satisfactory
signal quality (and, consequently, acceptable outage)
even when considering small antenna factors or omnidirectional transmissions.
In the second case, narrower beams
are needed to guarantee higher gains, produced by beamforming. 

Moreover, we observe that IEEE 802.11p systems usually provide more reliable communications than mmWave links since they present a lower outage probability.
Nevertheless, mmWave transmissions also achieve sufficient detection performance for spatially close vehicles (i.e., $d<110$~m) employing very narrow beams (e.g., $N = 64$).

Finally, for very large distances (i.e., $d>300$ m), all the investigated configurations achieve unacceptable reliability values. However, mmWave communications with sharp beams (e.g., $N>64$) have the potential to support unreliable long-distance inter-vehicle communications for which the IEEE 802.11p signal is basically undetectable.\\



\textbf{Robustness.}
The misalignment between the transmitter and the receiver, which can occur during lane change operations or as a consequence of the dissemination of inaccurate vehicle position information, may have a very detrimental impact on the performance of some \gls{v2v} applications, as sensors may provide skewed or delayed readings and vehicles may lose~connectivity.
In our scenario,  the degree $\delta_m$ of misalignment is distance-dependent and is given by $\delta_m=\arctan(W/d)$, where $W=2w=7$ m is the width of one carriageway lane and $d$ is the inter-vehicle distance.

 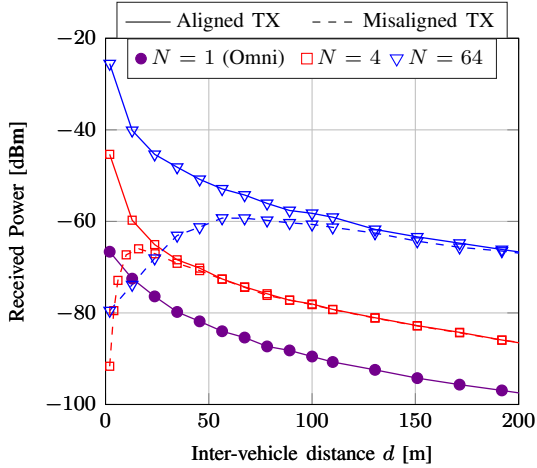
\begin{figure}[t!]
  \centering
	\setlength\fwidth{0.62\columnwidth}
	\setlength\fheight{0.55\columnwidth}
%
%
\definecolor{mycolor1}{rgb}{1.00000,0.00000,1.00000}%
\pgfplotsset{
tick label style={font=\footnotesize},
label style={font=\footnotesize},
legend  style={font=\footnotesize}
}
\begin{tikzpicture}

\begin{axis}[
width=\fwidth,
height=\fheight,
at={(0\fwidth,0\fheight)},
scale only axis,
xmin=0,
xmax=200,
xlabel style={font=\color{white!15!black}},
xlabel={VNs distance [m]},
ymin=-100,
ymax=-20,
ylabel style={font=\color{white!15!black}},
ylabel={Received Power [dBm]},
axis background/.style={fill=white},
xmajorgrids,
ymajorgrids,
xlabel={Inter-vehicle distance $d$ [m]},
label style={font=\footnotesize},
legend columns={2},
legend style={at={(0.5,1.1)},legend cell align=left, align=left,anchor=north, draw=white!15!black, draw=white!15!black,/tikz/every even column/.append style={column sep=0.35cm}},
]
\addplot [color = black]
  table[row sep=crcr]{%
0	0\\
};
\addlegendentry{Aligned TX}

\addplot [color = black,dashed]
  table[row sep=crcr]{%
0	0\\
};
\addlegendentry{Misaligned TX}

\end{axis}

\begin{axis}[%
width=\fwidth,
height=\fheight,
at={(0\fwidth,0\fheight)},
scale only axis,
xmin=0,
xmax=200,
xlabel style={font=\color{white!15!black}},
xlabel={VNs distance [m]},
ymin=-100,
ymax=-20,
ylabel style={font=\color{white!15!black}},
ylabel={Received Power [dBm]},
axis background/.style={fill=white},
xmajorgrids,
ymajorgrids,
xlabel={Inter-vehicle distance $d$ [m]},
label style={font=\footnotesize},
legend columns={-1},
legend style={at={(0.5,1)},legend cell align=left, align=left,anchor=north, draw=white!15!black,/tikz/every even column/.append style={column sep=0.05cm}}
]

\addplot [color=purple!60!blue,line width=0.5pt, mark size=2.0pt, mark=*, mark options={solid, rotate=180, purple!60!blue},forget plot]
  table[row sep=crcr]{%
2	-66.6283322203457\\
12.8888888888889	-72.5086669322114\\
23.7777777777778	-76.3862660598935\\
34.6666666666667	-79.7863301059444\\
45.5555555555556	-81.8399691358311\\
56.4444444444444	-84.0026519859774\\
67.3333333333333	-85.378996401462\\
78.2222222222222	-87.3206005768302\\
89.1111111111111	-88.1950758651659\\
100	-89.5216329048688\\
110	-90.7222068569828\\
130.473684210526	-92.438783104892\\
150.947368421053	-94.2454190461113\\
171.421052631579	-95.6663600230661\\
191.894736842105	-96.9313284451296\\
212.368421052632	-98.2774254321599\\
232.842105263158	-99.6550269425483\\
253.315789473684	-100.730230802802\\
273.789473684211	-101.899915828735\\
294.263157894737	-102.591897062675\\
314.736842105263	-103.766053619541\\
335.210526315789	-104.601163536884\\
355.684210526316	-105.4683288889\\
376.157894736842	-106.225442958209\\
396.631578947368	-106.69293132381\\
417.105263157895	-107.91238300097\\
437.578947368421	-108.495347546965\\
458.052631578947	-109.090099976673\\
478.526315789474	-109.637953583419\\
499	-110.435950068184\\
};

\addplot [color=red,line width=0.5pt, mark size=1.5pt, mark=square, mark options={solid, rotate=180, red},forget plot]
  table[row sep=crcr]{%
2	-45.3484430544428\\
12.8888888888889	-59.737781857263\\
23.7777777777778	-65.0721357938351\\
34.6666666666667	-68.3893070655338\\
45.5555555555556	-70.209065639991\\
56.4444444444444	-72.5710052844157\\
67.3333333333333	-74.3126926021003\\
78.2222222222222	-75.808093732626\\
89.1111111111111	-77.2289687642064\\
100	-78.0445890658198\\
110	-79.2239641170635\\
130.473684210526	-81.0473863725159\\
150.947368421053	-82.7760929496327\\
171.421052631579	-84.2383642073669\\
191.894736842105	-85.9191627933547\\
212.368421052632	-87.412129879439\\
232.842105263158	-88.2139368359412\\
253.315789473684	-89.5569537157655\\
273.789473684211	-90.6615288137715\\
294.263157894737	-91.5997980587943\\
314.736842105263	-92.3438199860634\\
335.210526315789	-93.2282532346778\\
355.684210526316	-94.2421547948765\\
376.157894736842	-95.1966286825098\\
396.631578947368	-95.7268271987022\\
417.105263157895	-96.7328939838561\\
437.578947368421	-97.3014391185403\\
458.052631578947	-98.1976253853997\\
478.526315789474	-98.5071832286769\\
499	-99.2076263097703\\
};

\addplot [color=red, dashed,line width=0.5pt, mark size=1.5pt, mark=square, mark options={solid, rotate=180, red},forget plot]
  table[row sep=crcr]{%
2	-91.650316952074\\
4	-79.4984606668088\\
6	-72.8898357450521\\
10	-67.3060824074549\\
16	-65.9837652452994\\
24	-67.0714309692721\\
34.6666666666667	-69.1525762593706\\
45.5555555555556	-70.7480003721683\\
56.4444444444444	-72.6483233698745\\
67.3333333333333	-74.411205969517\\
78.2222222222222	-76.1716968172336\\
89.1111111111111	-77.1915222290732\\
100	-78.1836812057127\\
110	-79.2779404833173\\
130.473684210526	-81.1327560453848\\
150.947368421053	-82.8393846496762\\
171.421052631579	-84.3395105304464\\
191.894736842105	-85.9595315357871\\
212.368421052632	-87.0440060312177\\
232.842105263158	-88.6558813498866\\
253.315789473684	-89.4041634926105\\
273.789473684211	-90.6090789896551\\
294.263157894737	-91.4813585664954\\
314.736842105263	-92.3442569001351\\
335.210526315789	-93.1629404555352\\
355.684210526316	-94.2305929617765\\
376.157894736842	-94.9601483144441\\
396.631578947368	-96.2010446605457\\
417.105263157895	-96.2146999340639\\
437.578947368421	-97.0241575212975\\
458.052631578947	-97.9065657520382\\
478.526315789474	-98.5658420167054\\
499	-99.3574013102388\\
};

\addplot [color=blue,line width=0.5pt, mark size=2.5pt, mark=triangle, mark options={solid, rotate=180, blue},forget plot]
  table[row sep=crcr]{%
2	-25.5334860045857\\
12.8888888888889	-40.1181924115973\\
23.7777777777778	-45.3513070269375\\
34.6666666666667	-48.1543314799351\\
45.5555555555556	-50.8093348831876\\
56.4444444444444	-52.8943220483104\\
67.3333333333333	-54.2363811911099\\
78.2222222222222	-56.0679743943892\\
89.1111111111111	-57.6169464199641\\
100	-58.2348429178112\\
110	-59.0847894960109\\
130.473684210526	-61.7469361611066\\
150.947368421053	-63.3826858744994\\
171.421052631579	-64.7584092076758\\
191.894736842105	-66.1376069301256\\
212.368421052632	-67.4586129952057\\
232.842105263158	-68.8192617679168\\
253.315789473684	-69.77577582146\\
273.789473684211	-70.6624999437177\\
294.263157894737	-71.5262261362236\\
314.736842105263	-72.7550265148143\\
335.210526315789	-73.8907402181952\\
355.684210526316	-74.2689013808285\\
376.157894736842	-75.3382649039444\\
396.631578947368	-75.7354080566739\\
417.105263157895	-76.5411377752621\\
437.578947368421	-77.5328859352076\\
458.052631578947	-77.9166035439902\\
478.526315789474	-78.7084013141314\\
499	-79.3596256328478\\
};

\addplot [color=blue, dashed,line width=0.5pt, mark size=2.5pt, mark=triangle, mark options={solid, rotate=180, blue},forget plot]
  table[row sep=crcr]{%
2	 -79.4978910169635\\
12.8888888888889	-73.8326418817044\\
23.7777777777778	-68.0581933026127\\
34.6666666666667	-63.0902003778139\\
45.5555555555556	-61.2511959847326\\
56.4444444444444	-59.280526909974\\
67.3333333333333	-59.3049377404534\\
78.2222222222222	-59.7745705367069\\
89.1111111111111	-60.3219024508525\\
100	-60.6532277586472\\
110	-61.2963841318194\\
130.473684210526	-62.5506026295239\\
150.947368421053	-64.3189628232185\\
171.421052631579	-65.6662583159722\\
191.894736842105	-66.4768301029492\\
212.368421052632	-67.5926918574598\\
232.842105263158	-69.2520729717707\\
253.315789473684	-70.2765169970806\\
273.789473684211	-71.0518618746208\\
294.263157894737	-72.161145303252\\
314.736842105263	-72.8595115554966\\
335.210526315789	-73.9895466645379\\
355.684210526316	-74.8561171085425\\
376.157894736842	-75.4984952231524\\
396.631578947368	-76.3147272576553\\
417.105263157895	-76.2851729506352\\
437.578947368421	-77.6200339191752\\
458.052631578947	-78.1859666468072\\
478.526315789474	-78.5436654954713\\
499	-79.7138462221173\\
};

\addplot [only marks,  mark size=2.0pt, mark=*, mark options={solid, rotate=180, purple!60!blue}]
  table[row sep=crcr]{%
0	0\\
};
\addlegendentry{$N=1$ (Omni) }

\addplot [only marks, mark size=2.0pt, mark=square, mark options={solid, rotate=180, red}]
  table[row sep=crcr]{%
0	0\\
};
\addlegendentry{$N=4$}

\addplot [only marks, mark size=2.8pt, mark=triangle, mark options={solid, rotate=180, blue}]
  table[row sep=crcr]{%
0	0\\
};
\addlegendentry{$N=64$}

\end{axis}

\end{tikzpicture}%
\vspace{-0.1cm}
    \caption{\footnotesize  Average received power \emph{vs} $d$ for aligned (solid lines) and misaligned (dashed lines) \gls{mmwave} communications, with different antenna array configurations (markers). }
    \label{fig:robustness_mmw}
  \end{figure}
  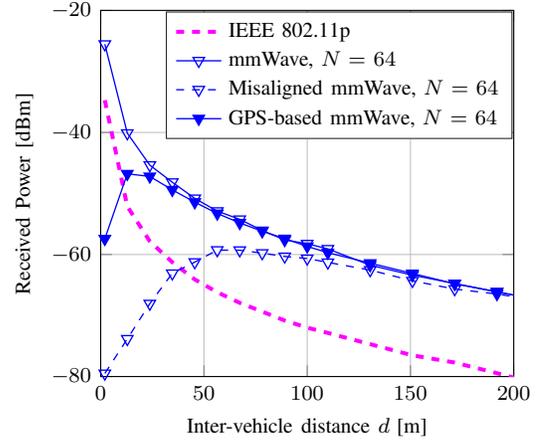
\begin{figure}[t!]
  \centering
	\setlength\fwidth{0.62\columnwidth}
	\setlength\fheight{0.55\columnwidth}
%
%

\definecolor{mycolor1}{rgb}{1.00000,0.00000,1.00000}%
\pgfplotsset{
tick label style={font=\footnotesize},
label style={font=\footnotesize},
legend  style={font=\footnotesize}
}
\begin{tikzpicture}

\begin{axis}[%
width=\fwidth,
height=\fheight,
at={(0\fwidth,0\fheight)},
scale only axis,
xmin=0,
xmax=200,
xlabel style={font=\color{white!15!black}},
xlabel={VNs distance [m]},
ymin=-80,
ymax=-20,
ylabel style={font=\color{white!15!black}},
ylabel={Received Power [dBm]},
axis background/.style={fill=white},
xmajorgrids,
ymajorgrids,
xlabel={Inter-vehicle distance $d$ [m]},
label style={font=\footnotesize},
legend columns={1},
legend style={at={(0.997,0.997)},legend cell align=left, align=center, draw=white!15!black},
]

\addplot [color=mycolor1, dashed, line width=1.5pt]
  table[row sep=crcr]{%
2	-34.713356311383\\
12.8888888888889	-52.1135110611529\\
23.7777777777778	-57.8398198902921\\
34.6666666666667	-61.2657711310337\\
45.5555555555556	-64.0910316634052\\
56.4444444444444	-66.1875713938211\\
67.3333333333333	-67.9290334580636\\
78.2222222222222	-69.3630789499177\\
89.1111111111111	-70.8273517358384\\
100	-71.9906843286553\\
110	-72.8562893761783\\
130.473684210526	-74.6944508573401\\
150.947368421053	-76.5306383424592\\
171.421052631579	-77.7109513403021\\
191.894736842105	-79.4533812623036\\
212.368421052632	-81.2766824376055\\
232.842105263158	-83.1514357454084\\
253.315789473684	-84.4697965164298\\
273.789473684211	-85.6662763976267\\
294.263157894737	-87.0529055545116\\
314.736842105263	-88.4418063388532\\
335.210526315789	-89.5857305087845\\
355.684210526316	-90.7615578713604\\
376.157894736842	-91.6749554298776\\
396.631578947368	-92.5292161178501\\
417.105263157895	-93.3316569911229\\
437.578947368421	-94.3011873518775\\
458.052631578947	-94.511565464579\\
478.526315789474	-95.7206685750302\\
499	-96.2046676616603\\
};
\addlegendentry{IEEE 802.11p}

\addplot [color=blue,line width=0.5pt, mark size=2.5pt, mark=triangle, mark options={solid, rotate=180, blue}]
  table[row sep=crcr]{%
2	-25.5334860045857\\
12.8888888888889	-40.1181924115973\\
23.7777777777778	-45.3513070269375\\
34.6666666666667	-48.1543314799351\\
45.5555555555556	-50.8093348831876\\
56.4444444444444	-52.8943220483104\\
67.3333333333333	-54.2363811911099\\
78.2222222222222	-56.0679743943892\\
89.1111111111111	-57.6169464199641\\
100	-58.2348429178112\\
110	-59.0847894960109\\
130.473684210526	-61.7469361611066\\
150.947368421053	-63.3826858744994\\
171.421052631579	-64.7584092076758\\
191.894736842105	-66.1376069301256\\
212.368421052632	-67.4586129952057\\
232.842105263158	-68.8192617679168\\
253.315789473684	-69.77577582146\\
273.789473684211	-70.6624999437177\\
294.263157894737	-71.5262261362236\\
314.736842105263	-72.7550265148143\\
335.210526315789	-73.8907402181952\\
355.684210526316	-74.2689013808285\\
376.157894736842	-75.3382649039444\\
396.631578947368	-75.7354080566739\\
417.105263157895	-76.5411377752621\\
437.578947368421	-77.5328859352076\\
458.052631578947	-77.9166035439902\\
478.526315789474	-78.7084013141314\\
499	-79.3596256328478\\
};
\addlegendentry{mmWave, $N = 64$}

\addplot [color=blue, dashed,line width=0.5pt, mark size=2.5pt, mark=triangle, mark options={solid, rotate=180, blue}]
  table[row sep=crcr]{%
2	 -79.4978910169635\\
12.8888888888889	-73.8326418817044\\
23.7777777777778	-68.0581933026127\\
34.6666666666667	-63.0902003778139\\
45.5555555555556	-61.2511959847326\\
56.4444444444444	-59.280526909974\\
67.3333333333333	-59.3049377404534\\
78.2222222222222	-59.7745705367069\\
89.1111111111111	-60.3219024508525\\
100	-60.6532277586472\\
110	-61.2963841318194\\
130.473684210526	-62.5506026295239\\
150.947368421053	-64.3189628232185\\
171.421052631579	-65.6662583159722\\
191.894736842105	-66.4768301029492\\
212.368421052632	-67.5926918574598\\
232.842105263158	-69.2520729717707\\
253.315789473684	-70.2765169970806\\
273.789473684211	-71.0518618746208\\
294.263157894737	-72.161145303252\\
314.736842105263	-72.8595115554966\\
335.210526315789	-73.9895466645379\\
355.684210526316	-74.8561171085425\\
376.157894736842	-75.4984952231524\\
396.631578947368	-76.3147272576553\\
417.105263157895	-76.2851729506352\\
437.578947368421	-77.6200339191752\\
458.052631578947	-78.1859666468072\\
478.526315789474	-78.5436654954713\\
499	-79.7138462221173\\
};
\addlegendentry{Misaligned mmWave, $N=64$}

\addplot [color=blue, line width=0.5pt, mark size=2.5pt, mark=triangle*, mark options={solid, rotate=180, blue}]
  table[row sep=crcr]{%
2	-57.4301575424184\\
12.8888888888889	-46.7783851693804\\
23.7777777777778	-47.2082327144499\\
34.6666666666667	-49.4030352977458\\
45.5555555555556	-51.424373310255\\
56.4444444444444	-53.2808654335917\\
67.3333333333333	-54.8204831901733\\
78.2222222222222	-56.1899164424784\\
89.1111111111111	-57.3934971399096\\
100	-58.6726468913115\\
110	-59.7035817471034\\
130.473684210526	-61.4170791247584\\
150.947368421053	-63.1076011757326\\
171.421052631579	-64.799175449135\\
191.894736842105	-66.1169742709199\\
212.368421052632	-67.4731708727238\\
232.842105263158	-68.6450661456699\\
253.315789473684	-69.589074276651\\
273.789473684211	-70.6719780175386\\
294.263157894737	-71.6343310923547\\
314.736842105263	-72.67203042003\\
335.210526315789	-73.6450949178114\\
355.684210526316	-74.4003128929619\\
376.157894736842	-75.1134773405481\\
396.631578947368	-75.9290698915284\\
417.105263157895	-76.593458526772\\
437.578947368421	-77.4010810034829\\
458.052631578947	-78.119159539794\\
478.526315789474	-78.847614214831\\
499	-79.3035611395514\\
};
\addlegendentry{GPS-based mmWave, $N = 64$}

\end{axis}
\end{tikzpicture}%
    \caption{\footnotesize  Average received power \emph{vs} $d$ for IEEE 802.11p (dashed line) and directional \gls{mmwave} communications (solid lines), with $N=64$. If applicable, the inaccuracy of the geographical position data is modeled according to a Gamma distribution with parameters $\alpha = 3.14733$ and $\beta = 0.462432$~\cite{driver2007long}. }
    \label{fig:robustness_dir}
  \end{figure}

In Fig.~\ref{fig:robustness_mmw}, we observe that the impact of the misalignment on the communication performance depends on several factors, including  $d$  and  the beamwidth.
In case of very directional \gls{mmwave} transmissions (e.g., $N=64$), the quality of the received signal significantly decreases as a result of misaligned nodes (i.e.,  more than 50 dB for short distances), mainly due to the non-continuous beamtracking mechanism: after the alignment is lost, vehicles need to wait for a new tracking operation to be performed to recover their optimal beam configuration.
Conversely, more robust alignment can be achieved when considering smaller~array factors  since wider beams enlarge the area in which the vehicles are within coverage. 
Omnidirectional strategies are not  affected by misalignment (in Fig.~\ref{fig:robustness_mmw}, the solid and dashed curves for the $N=1$ case  overlap perfectly).
In this approach, however,  the well-known  robustness versus throughput trade-off is exposed  \cite{giordani2018coverage}:  wider beams guarantee more robust and continuous connectivity but generally yield lower received power and transmission rates, as shown in~Fig.~\ref{fig:datarate}.

Moreover, from Fig.~\ref{fig:robustness_mmw} we see that the impact of misalignment is dominant at short ranges. Indeed, the received power initially increases with $d$ since, for larger distances, the projection of the beam’s shape onto the road surface is geometrically larger,  thereby increasing the maximum distance that the vehicles can cover before leaving their respective communication ranges.
However, beyond a certain value of $d$ (i.e., $d>50$~m for $N=64$ and $d>10$~m for $N=4$), beams are already sufficiently large to allow for loose alignment and the received power decreases just because of the path loss.


If sensory information (e.g., GPS coordinates) is available, it can be used to  aid the configuration of the \gls{mmwave} communication link and to remove the need  for periodical  beam  tracking  operations.
In this regard,   Fig.~\ref{fig:robustness_dir} reports the effect of misalignment due to  inaccurate data, which makes the nodes point their beams towards improper directions. 
Nevertheless,  such  inaccuracy compromises only very short-range transmissions (i.e., $d<20$ m for $N=64$).\footnote{ The accuracy of the position information may be improved by the adoption of \emph{data fusion} strategies which combine several localization techniques, e.g., dead reckoning, cellular localization, and camera image processing, into a single solution that is more robust and precise than any individual approach~\cite{alam2013relative}.}

Finally,  Fig.~\ref{fig:robustness_dir} exemplifies how the omnidirectional transmissions of  IEEE 802.11p  systems offer more robust and, in some circumstances (e.g., $d<40$), more efficient \gls{v2v} communications than their \gls{mmwave}~counterparts.

\section{Discussion and Future Work} 
\label{sec:conclusions_and_future_work}

In this paper we provided the first numerical comparison of the performance, in terms of achievable data rate, detection accuracy and robustness, between  IEEE 802.11p (the current standard for short-range vehicular communications) and the mmWave technology  to support \gls{v2v} networking. 
Overall, we showed that IEEE 802.11p systems  offer low-rate connectivity (i.e., up to a few tens of Mbps) but guarantee very stable, reliable and robust transmissions at short/medium distances (i.e., up to a few hundreds of meters) thanks to the intrinsic stability of the low-frequency channels and the omnidirectional transmissions.
Conversely,  \gls{mmwave} systems support very high-throughput connections but  exhibit high instability due to the severe signal propagation characteristics and the need to maintain beam alignment.
Although the connectivity robustness can be magnified by considering transmissions through wide beams, the data rate increases  considering very directional communications and close-range vehicles instead.
We also showed that sensory information has the potential to  help  reducing the beam alignment overhead  with minor performance degradation even in the presence of inaccurate~data.

In this context,  we conclude that the synergistic orchestration among the different radios makes it possible to complement the limitations of each type of network and deliver more flexible and resilient transmissions.

This work opens up  some particularly interesting research directions, such as the definition of  interface selection mechanisms able to dynamically  identify the recommended type of radio  to interconnect the vehicles.
Moreover, due to the lack of temporally and spatially correlated channel measurements in the mmWave band, it is currently not possible to accurately evaluate the performance of the vehicles in realistic mobility-related scenarios, which on the other hand remains a very interesting and relevant item for future research.



\bibliographystyle{IEEEtran}
\bibliography{bibliography.bib}

\end{document}